\documentclass[%draft % uncomment if draft mode is needed
               ]{jetp} % jetp.cls
\twocolumn % закомментировать, если не нужен двухколоночный вариант
\begin{document}

\title{Black Hole Shadows Constrain Extended Gravity}% Force line breaks with \\

%%\thanks{A footnote to the article title}%

\setaffiliation1{Sternberg Astronomical Institute, Lomonosov Moscow State University, Moscow 119992, Russian Federation.
}

\setaffiliation2{
Department of Quantum Theory and High Energy Physics, Physics Faculty, Lomonosov Moscow State University, Moscow 119992, Russian Federation.
}%

\setaffiliation3{Department of Astrophysics and Stellar Astronomy, Physics Faculty, Lomonosov Moscow State University, Moscow 119992, Russian Federation.}

\setauthor{Vjacheslav}{Prokopov}{13}
\setauthor{Stanislav}{Alexeyev}{12}
\setauthor{Oleg}{Zenin}{2}

\date{\today}% It is always \today, today,

\abstract{The first images of black hole shadows open new possibilities to constrain modern extended gravity theories. We present the method of shadow background calculation for black hole solutions in the form of Taylor series where $g_{11} = - g_{00}^{-1}$. The method is extended to general non-rotating case $g_{11} \neq - g_{00}^{-1}$. The results of the analysis are compared with the predictions of General relativity  taking into account the Event Horizon Telescope data. The results for the Horndesky model with the Gauss-Bonnet invariant, loop quantum gravity, Bumblebee model and Gauss-Bonnet gravity are in full agreement with the observations of M87*. In conformal gravity, large values of $m_2$ and $Q_s$ must be excluded. In STEGR $f(Q)$ gravity the observational limits on the parameter $\alpha$ are: $-0.025<\alpha<0.04$. For an alternative generalization of the Bumblebee model with the Schwarzschild approximation: $-0.3 < l < 0.45$. These results demonstrate the maximum one can achieve without taking into account of the rotation of a black hole.}

\maketitle

%\tableofcontents

\section{Introduction}

First spherically-symmetric solutions were discovered more than 100 years ago. The existence of real objects described by these spherically-symmetric and axially-symmetric metrics was proven by the observations not so long ago. The results on binary systems dynamics \cite{Orosz:2011np}, gravitational wave astronomy \cite{TheLIGOScientific:2016qqj, Abbott_2017}, direct imaging of black hole (BH) shadows \cite{Akiyama:2019fyp} are the most well-known examples. Currently the General Theory of Relativity (GR) reproduces the astronomical data with great accuracy. Meanwhile such problems as dark matter, dark energy, the early Universe evolution, quantum gravity, ... are waiting for a better theoretical basis. So, new extended gravity models are developed to explain these phenomena better. They are $f(R)$ gravity \cite{Capozziello:2011et}, $f(Q)$ gravity \cite{DAmbrosio:2021zpm}, scalar-tensor theories including the most general case with second order field equations: Horndesky theory \cite{Sotiriou:2008rp, DeFelice:2010aj, Charmousis:2011bf, Babichev:2017guv}, teleparallel models \cite{Pfeifer:2021njm}, gravity models with conformal symmetry \cite{Mannheim_2011,Myung_2019}, loop quantum gravity \cite{casares2018review,De_Lorenzo_2015,Hu_2018}, scalar Gauss–Bonnet gravity \cite{PhysRevD.83.104002} and other approaches. It seems important to constrain these extended gravity theories. So, the achievements in black hole shadow imaging provide additional possibility for this. 

Let us briefly remind the key feathers of the models discussed further. We start from the Horndesky model \cite{Horndeski:1974wa}. It represents the most general case of scalar-tensor gravity producing second order field equations \cite{Kobayashi:2019hrl}. Horndesky model could model dark energy or dark matter. It seems to be more fundamental than pure Brance-Dicke model. After GW170817 Horndesky theory was severely limited. Now it is used in the form of DHOST (degenerated higher-order scalar-tensor) theories \cite{2021ag}. Further Horndesky theory often is combined with the Gauss-Bonnet invariant $S_{GB} = R_{\alpha\beta\gamma\sigma} R^{\alpha\beta\gamma\sigma} - 4 R_{\alpha\beta} R^{\alpha\beta} + R^2$ (\cite{Babichev:2017guv}) where $R_{\alpha\beta\gamma\sigma}$, $R_{\alpha\beta}$ and $R$ are curvature tensors and scalar. 

The next model under consideration is loop quantum gravity (LQG). LQG represents a perspective approach to construct a quantum theory of gravity. The key idea is the independence of other physical interactions and the application of the specific choice of parametric space. Theory functions form the closed algebra of operators allowing to construct a renormalizable theory. LQG allows to combine bounce and inflation stages and to reproduce the theory of early Universe \cite{Barrau:2013ula}. 

Going further we mention gravity models with conformal symmetry \cite{Mannheim:2011ds}. Such a symmetry in the action gives perspectives to construct a renormalizable theory of gravity. Linear realizations have fourth order field equations. Now the community considers models with nonlinear symmetry realization \cite{Arbuzov:2019rcl, Alexeyev:2020lag, Alexeyev:2021lav}). These models are developed in a short time. They have a set of problems. For example, there is no inflation asymptotic \cite{Alexeyev:2021lav}. If these problems would solved they seems to be perspective (in addition to quantum gravity) in dark energy.  

The next example is Bumblebee model. This model extends the standard GR by a vector field. Under a suitable potential the Bumblebee vector field  $B_\mu$ acquires a nonvanishing vacuum expectation value. Such combination induces a spontaneous Lorentz symmetry breaking \cite{2018}. The discussed approach could form a ``bridge'' between the string theory at Planckian scales and GR solving GR problems in the high energy range.

Next we discuss the Teleparralel Equivalent of General Relativity (TEGR). Here GR is considered with non-vanishing torsion and non-metricity. Therefore geometrical deformation causes gravitational field directly. TEGR allows to include additional degrees of freedom to describe GR non-solved phenomenons. We consider $f(Q)$ gravity which is a symmetric TEGR (STEGR) where the non-metricity scalar $Q$ is not equal to zero.\cite{DAmbrosio:2021zpm}. 

The last model is Scalar Gauss-Bonnet gravity. This is a modified theory with actions including all possible quadratic curvature scalars \cite{PhysRevD.83.104002}. The curvature scalars play the same role as in the previous case. Being a phenomenological asymptote of some geometry they could provide physical explanation of GR unresolved problems.  

From the other side the real physical equipment has the limited accuracy. Therefore each experimental result admits few alternative explanations caused by different theories \cite{Lima:2021las}. At the first step usually the most simple model is for this chosen. Further additional data allows to narrow down the set of alternative explanations. So the shadow size value being the first one measured at observations could be applied for the additional estimation of the model predictions. Therefore we use the standard GR space-times (Schwarzschild, Kerr, ...) as basic approximations.

Previously we discussed a shadow form and size, last stable orbit and strong gravitational lensing calculations when the third approximation in spherically-symmetric space-time is taken into account. Such metrics represent the continuation of Reissner-Nordstrom space-time by the next expansion order relatively $r^{-1}$ (\cite{Alexeyev:2019keq}. Further when the rotation had been included, the shape of the shadow became sufficient to test theories beyond Kerr-Newman space-time\cite{Alexeyev:2020frp}. So to define the expansion coefficients one has to use the observational results of shadow size, last stable orbit and strong gravitational lensing. For example, when the third correction is under consideration two different probes are required to restore BH characteristics. For calculation of next expansion orders one has to increase the amount of probes. 

In this paper we discuss how to constrain some extended gravity models using modern data on BH images. Here it is necessary to point out that in new gravity models the activity firstly is concentrated on non-rotating BH solutions as more simple ones. Therefore we develop the formalism for spherically-symmetric space-time to extract maximum information in lighter case (as a first step of general study). Our consideration is extended to the case $g_{11} \neq - g_{00}^{-1}$. It was shown \cite{2021} that the maximum variation of a shadow size for a rotating BH amounts about $5-7\%$. In the spin value is small the influence of a rotation on the shadow size can be neglected. The first limits on the BH shadow size in M87* observation were obtained in \cite{2021}. They are: $4.31M < D < 6.08M$.

The paper is organized as follows. Section 2 is devoted to the degenerated case $g_{11} = - g_{00}^{-1}$, Section 3 extends the consideration to $g_{11} \neq - g_{00}^{-1}$ case, at Section 4 we discuss the examples for gravity models mentioned above and Section 5 contains our conclusions.

\section{Shadow model at $A (r) =  B^{-1}(r)$.}

The general description of asymptotically flat static spherically-symmetric space-time in modified gravity represents the extension of Schwarzschild metric in the form:
\begin{eqnarray}\label{f1}
ds^2 =  -  A(r) dt^2 + B(r) dr^2 + r^2 (d\theta^2+ \sin^2\theta d\phi^2 ),
\end{eqnarray}
where $A(r)$ and $B(r)$ are metric functions depending upon radial coordinate $r$. The standard Schwarzschild metrics in Planckian units $G=c=\hbar=1$ has the form 
\begin{eqnarray}\label{f2}
A(r) =  B^{-1}(r) = 1 - \frac{2M}{r} , 
\end{eqnarray}
where $M$ is BH mass. Note that Schwarzschild metric, Reissner-Nordstrom one and further extensions represent the terms of Taylor expansion when $r >> 2M$. In this approach one considers the Schwarzschild metric as a first approximation. In this approximation one can describe the star's trajectories around central BH. Reissner-Nordstrom metric as next expansion order allows to describe the influence of electrical or tidal charges \cite{Dadhich:2000am}. Appearance of a tidal charge sometimes drastically changes shadow properties \cite{Zakharov:2014lqa, Pugliese:2010ps, Alexeyev:2019keq, Alexeyev:2020frp}.  

We start from a degenerated case of ``symmetrical'' metric functions $A(r) = B(r)^{-1}$ in Eq.(\ref{f1}). So the event horizon position is defined as $A(r)=0$. When the solution of this equation is not unique one has to consider the external one. At the next step one restricts the series of $A(r)$ by the third expansion order as:
\begin{eqnarray}\label{rnm+}
A(r) = 1 - \cfrac{2M}{r} + \cfrac{Q}{r^2} + \cfrac{C_3}{r^3}, 
\end{eqnarray} 
where $Q$ is a tidal charge and $C_3$ is a Taylor expansion coefficient at $r^{-3}$ order. To simplify the calculations one can normalize all the values on BH mass: $\hat{r}=r/M$, $q=Q/M^2$ and $c_3=C_3/M^3$, therefore $\hat{r}$, $q$ and $c_3$ become unit free. Hence the configuration space appears to be two dimensional and:
\begin{eqnarray}\label{nm1}
A (\hat{r})=1 - \cfrac{2}{\hat{r}} + \cfrac{q}{\hat{r}^2} + \cfrac{c_3}{\hat{r}^3} ~.
\end{eqnarray}

The set of unstable photon orbits forms a photon sphere and, hence, defines the boundary of a BH shadow. Photons from a far distance source with sighting parameter $b$ greater than critical value $b_{ph}$ pass outside the sphere and, further, reach the external observer. Other photons with $b < b_{ph}$ interact with the BH and form a spot at the image, i.e. BH shadow. Hence the visible shadow image from non-rotating (or slowly rotating) BH has a form of a disk. Its radius is defined by the critical sighting parameter ($b_{ph} = 3\sqrt{3} M$ for Schwarzschild BH \cite{Bambi:2013nla}). The form of a shadow image may be distorted by a strong gravitational lensing. 

Consider the optically thin accretion disk surrounding the compact object \cite{Shakura:1976xk}. We follow \cite{Bambi:2013nla} and modify his approach for the symmetric case $A(r)=B(r)$ (Eq. (\ref{f1})). So the radiation is emitted from the surface situated outside the horizon including the regions inside the photon sphere. Therefore the specific intensity $I_{\nu_{0}}$ that could be measured (usually in $erg s^{-1} cm^{-2} str^{-2} Hz^{-1}$) at visible photon frequency  $\nu_{0}$ and the position $(X,Y)$ (coordinates at image plane) on the sky sphere is equal to \cite{Bambi:2013nla}:
\begin{eqnarray}\label{I1}
I_{\nu_{0}}=\int_{\gamma} z^3j(\nu_{e})dl_{prop}~,
\end{eqnarray}
where $\nu_{e}$ is emitted frequency, $z=\nu_{0}/\nu_{e}$ is redshift, $j(\nu_{e})$ is volume unit emitting potential of resting source, $dl_{prop}=-k_{\alpha}u^{\alpha}_{e}d\lambda$ is the differential of length unit in the source frame, $k^{\mu}$ is 4-speed of the photon, $u^{\mu}_{e}$ is 4-speed of the BH and $\lambda$ is the affine parameter along the photon $\gamma$ trajectory. Index $\gamma$ means the integration along isotropic geodesics. Red-shift $z$ is defined as \cite{Bambi:2013nla}:
\begin{eqnarray}\label{nm2}
z = \frac{k_{\alpha}u^{\alpha}_{0}}{k_{\beta}u^{\beta}_{e}}~,
\end{eqnarray}
where $u^{\mu}_{0}=(1,0,0,0)$ is 4-speed of a distance observer.

Considering the simple spherically-symmetric model of accretion we suppose that the gas freely falls in radial direction to the BH center with the following 4-speed:
\begin{eqnarray}\label{nm3}
u^t_{e}=\frac{1}{A(\hat{r})},    u^r_{e}=-\sqrt{1-A(\hat{r})},    u^{\theta}_{e}=u^{\phi}_{e}=0~.
\end{eqnarray}
$k^{\mu}=\dot{x}^{\mu}$ was calculated in \cite{Bambi:2013nla}, therefore, combining these results one obtains:
\begin{eqnarray}\label{nm4}
\frac{k_{r}}{k_{t}}=\pm \sqrt{\frac{1}{A(\hat{r})}[\frac{1}{A(\hat{r})}-\frac{b^2}{\hat{r}^2}]}~,
\end{eqnarray}
where the sign ``+(-)'' denotes the moving from (to) BH. So the redshift transfers to \cite{Shaikh_2018}:
\begin{eqnarray}\label{nm41}
z=\frac{1}{\frac{1}{A(\hat{r})}-\frac{k_{r}}{k_{t}}\sqrt{1-A(\hat{r})}}~.
\end{eqnarray}

For the shadow profile we suppose that the frequency of resting source is $\nu_{*}$, radiation is monochromatic and has the radial profile $1/\hat{r}^2$ \cite{Bambi:2013nla}:
\begin{eqnarray}\label{nm5}
j(\nu_{e})\propto\frac{\delta(\nu_{e}-\nu_{*})}{\hat{r}^2}~,
\end{eqnarray}
where $\delta$ is Dirac delta-function. The differential of length unit in the resting frame is defined as:
\begin{eqnarray}\label{nm6}
dl_{propo}=-k_{\alpha}u^{\alpha}_{e}d\lambda=-\frac{k^t}{zk^r}dr~.
\end{eqnarray}
Integrating the Eq. (\ref{I1}) over all the observed frequencies one obtains the observable intensity of photons at the position $(X,Y)$ on sky sphere \cite{Bambi:2013nla}:
\begin{eqnarray}\label{I3}
I_{obs}(X,Y)\propto\int_{\gamma} \frac{z^3k^t dr}{r^2 k^r}~.
\end{eqnarray}
The value of sighting parameter $b$ depends upon the position at the image plane $(X,Y)$ and is equal to $b^2\propto X^2+Y^2$. After the numerical integration we obtain the intensity profile of BH shadow. 

The dependence of shadow size upon $q$ and $c_3$ was calculated earlier \cite{Alexeyev:2019keq}. It was shown that if the shadow size is greater than $4M$ only one additional degree of freedom (namely $q$) is necessary. Hence in the first expansion order such a shadow can be parametrized by the Reissner-Nordstrom metric. Further when the intensity profile starts to change one has to incorporate next perturbation orders.

When the third and further expansions are taken into account the shadow description appears to be not unique. It allows a set of different parameter combinations because the increasing of equation's order causes the appearing of addition solutions. Hence more observational data would require to constrain the theoretical model. So in addition to the shadow size one has to consider the last stable orbit radius, strong gravitational lensing of the bright object close to BH and the distribution of background intensity. As previously the consideration starts from the Schwarzschild space-time.

\begin{figure}
\begin{center}
\includegraphics[width=\linewidth]{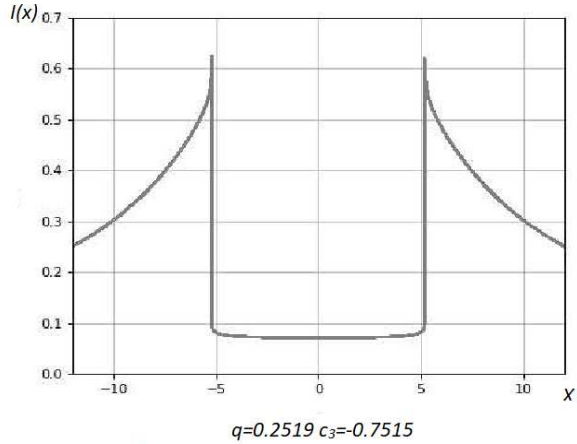}
\end{center}
\caption{The dependence of the intensity profile ($I$) in relative units for the shadow versus the distance from BH center on image plane $X$ (in the units of $M$). The BH possesses by additional parameters $q$ and $c_3$ and generates the shadow size the same as Schwarzschild BH of the equal size.}
\label{sh=}
\end{figure}

\begin{figure}
\begin{center}
\includegraphics[width=\linewidth]{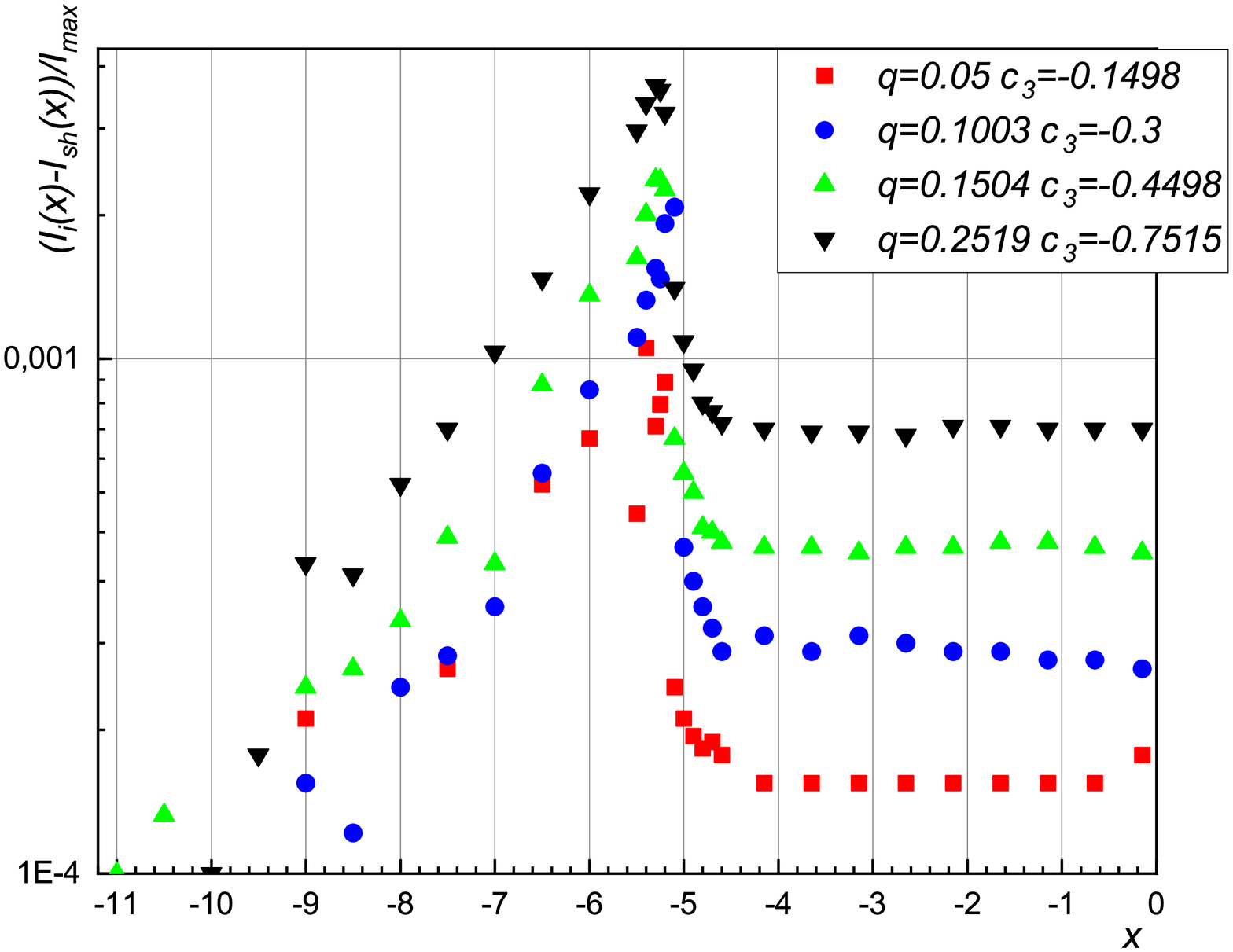}
\end{center}
\caption{The dependence of the intensity difference ($|I-I_{Sh}|/I_{max}$) between a BH with additional parameters $q$ and $c_3$ and the corresponding Schwarzschild one versus the distance from BH center on image plane $X$ in the units of $M$.}
\label{ressh}
\end{figure}

The figure \ref{sh=} demonstrates the intensity of the BH shadow profile  with  $q=0.2519$, $c_3=-0.7515$. The key moment is that its size is equal to the Schwarzschild BH one. The difference from the Schwarzschild BH (normalized on maximal intensity) is: $I_{max} \approx 0.6$ (Fig. \ref{ressh}). As one can see from Fig. \ref{ressh}, this difference grows when additional parameters increase. The maximal difference takes place near the shadow boundary, then it vanishes while going to the infinity. The difference inside shadow is constant. Further, from Fig. \ref{ressh} one concludes that the intensity resolution greater than $0.1\%$ of maximal intensity is required to fix this difference in observations. Note that each point of the profile could serve as an addition probe of BH potential.

\section{Shadow model at $A (r) \neq  B^{-1}(r)$.}

In general spherically-symmetric space-time $A (r) \neq  B^{-1}(r)$ (Eq. (\ref{f1})). To extend our consideration for this case we start from equations of motion in the form:
\begin{eqnarray} 
&& \left(\cfrac{d \hat{r}}{d\tau}  \right)^2 +  \cfrac{L^2}{B(\hat{r})\hat{r}^2}=\cfrac{E^2}{A(\hat{r})B(\hat{r})}, \label{fr} \\ 
&& \cfrac{d\phi}{d\tau} = \cfrac{L}{\hat{r}^2}, \label{ft} 
\end{eqnarray}
where $E$ is photon energy, $L$ is the angular momentum of the photon beam and $\tau$ is affine parameter. After substitution Eq. \eqref{ft} to Eq. \eqref{fr} these equations transform to:
\begin{eqnarray}\label{rf}
u(r) = \left(\cfrac{d \hat{r}}{d \phi}\right)^2 = \cfrac{\hat{r}^4}{D^2 A(\hat{r})B(\hat{r})} - \cfrac{ \hat{r}^2}{B(\hat{r})},
\end{eqnarray}
where $D = L/E$ is sighting parameter of the photon beam. Analogously to the symmetric case the shadow end occurs when the photon trajectory becomes unstable. Therefore the corresponding equation is:
\begin{eqnarray}\label{ss}
 u(r)=0 , & \cfrac{du(r)}{dr} = 0 , & \cfrac{d^2u(r)}{d^2r} > 0 .
\end{eqnarray}
To calculate the shadow size one has to find the maximal root of Eqs (\ref{ss}). We proceed this numerically.

\section{Shadows in Different Theories}

\subsection{Horndesky Theory} 

\begin{figure}                     
\begin{center}
\includegraphics[width=\linewidth]{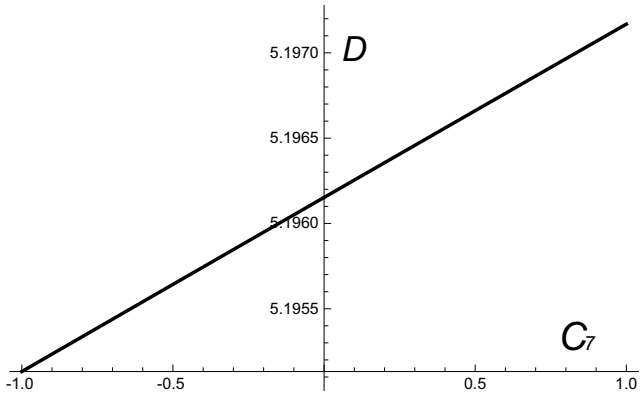}
\end{center}
\caption{The dependence of shadow size ($D$) versus the combination of model constants $C_7$ for Horndesky theory coupled with Gauss-Bonnet invariant (in the units of $M$, $M=1$).}
\label{r7}
\end{figure}

We consider the BH solution in Horndesky theory linearly coupled with Gauss-Bonnet invariant \cite{Babichev:2017guv}. So the metric functions from Eq.(\ref{f1}) are:
\begin{eqnarray}\label{f6}
A(r) & = & 1-\cfrac{2M}{r}-\cfrac{2C_7}{7r^7}\\
B(r)^{-1} & = & 1-\cfrac{2M}{r}-\cfrac{C_7}{r^7}
\end{eqnarray}
where $C_7$ is the specific combination of model constants. Only positive values of $C_7$ were considered in \cite{Babichev:2017guv}. Otherwise the requirement to the position of the horizon $A(r_h)=0$ to be located outside the surface $B(r)=0$ (to avoid singularities) fulfills only if $C_7<0$. So when $C_7 > 0$ this object is not a BH. Hence it is reasonable to suppose that the metric (\ref{f6}) is valid outside the photon sphere. Near the horizon a more accurate expansion (compatible with the BH definition) is required. The numerical results on shadow size dependence upon metric functions from Eq. (\ref{f6}) are presented at Fig. \ref{r7}. The shadow size for metric functions (\ref{f6}) differs from corresponding Schwarzschild values less than $0.01\%$ (at $|C_7|<0.5$). This occurs even when  $C_7$ value is compatible with $M$. Hence the observational data compatible with GR also does not forbid Horndesky theory.  

\subsection{Loop Quantum Gravity}

\begin{figure}                    
\begin{center}
\includegraphics[width=\linewidth]{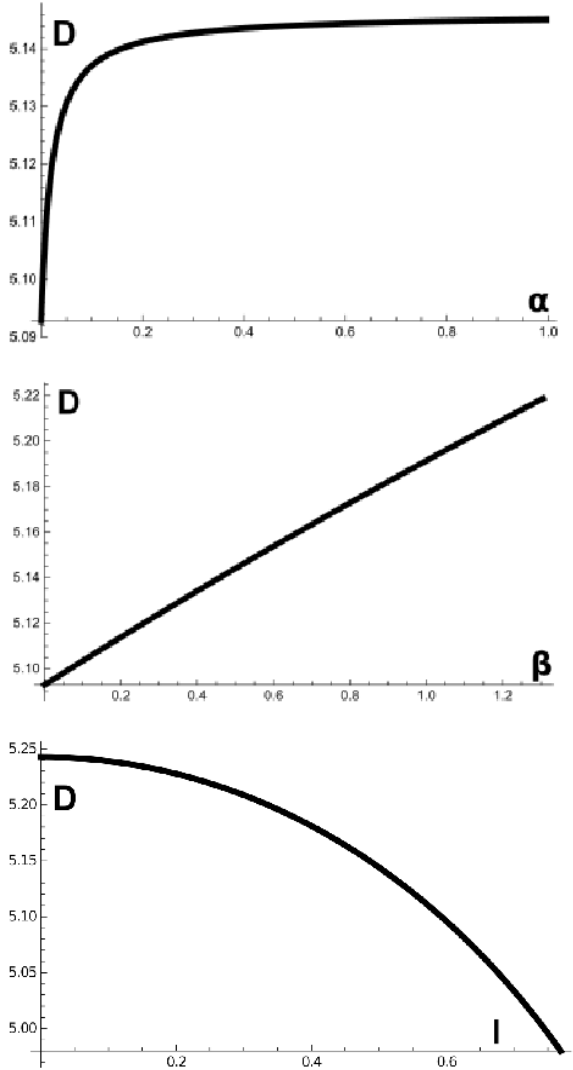}
\caption{The dependence of shadow size $D$ upon the time delay $\alpha$ when  $l=0.5M$ and $\beta=0.5$ (top image), upon the 1-loop quantum corrections $\beta$ when  $l=0.5M$, $\alpha=0.5$ (central image), upon the central energy density $l$ when  $\alpha=0.5$, $\beta=0.5$ (bottom image) for BH in modified Hayward metric in the units of $M$, $M=1$.}
\label{r8}
\end{center}
\end{figure}

As the next application we discuss LQG with the modified Hayward metric \cite{De_Lorenzo_2015,Hu_2018} as BH solution. This metrics has no central singularity. That is why it is called as ``regular BH''. Its extension includes a time delay and the 1-loop quantum correction. So the metric functions in this case are:
\begin{eqnarray}\label{f7}
A(r) & = & (1-\cfrac{2Mr^2}{r^3+2Ml^2})(1-\cfrac{\alpha\beta M}{\alpha r^3+ \beta M})\\
B(r)^{-1} & = & 1-\cfrac{2Mr^2}{r^3+2Ml^2}
\end{eqnarray}
where $l$ encodes the central energy density $3/8\pi l^2$, the constant $\alpha$ is the time delay between the center and infinity and $\beta$ is related to the 1-loop quantum corrections of the Newtonian potential. These parameters were constrained in \cite{De_Lorenzo_2015,Hu_2018} as: $0\leq \alpha<1$, $\beta_{max}=41/(10\pi)$. When $l >\sqrt{16/27}M$ the object has no horizon. After solving the Eqs. (\ref{f6}) one obtains the dependencies (presented at Fig. \ref{r8}) of the shadow size value upon $l$, $\alpha$ and $\beta$. One can see that during increasing of $l$ the shadow size decreases. In opposite the increasing of $\alpha$ and $\beta$ leads to shadow size increasing. When $\beta \geq 0$ the minimal shadow size is reached at $l=\sqrt{16/27}M$, $\beta=\alpha=0$ being equal to $4.92M$. The maximal shadow size occurs when $l=0$, $\beta=41/(10\pi)$, $\alpha=1$ and is equal to $5.32M$. Note that the shadows of such size could be described also by the Reissner-Nordstrom space-time. So using the shadow size only it is impossible to extract all parameters without additional observational data.

\subsection{Conformal Gravity} 

\begin{figure}                     
\begin{center}
\includegraphics[width=\linewidth]{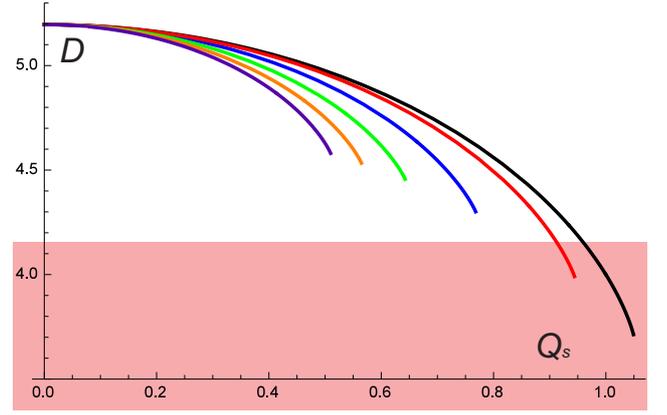}
\end{center}
\caption{The dependence of the shadow size $D$ against the scalar charge $Q_s$ for in new massive conformal gravity with different values of massive spin-2 mode $m_2$ (in the units of $M$, $M=1$). Black line corresponds to $m_2 \rightarrow \infty$, red one corresponds to $m_2=2$, blue one corresponds to $m_2=1$, green one corresponds to $m_2=0.707$, orange one corresponds to $m_2=0.577$, purple one corresponds to $m_2=0.5$.}
\label{r11}
\end{figure}

The next example is gravity model with conformal symmetry. As an example we choose BH metric in new massive conformal gravity \cite{Myung_2019}: 
\begin{eqnarray}\label{f8}
A(r) & = & 1-\cfrac{2M}{r}+\cfrac{Q_s^2}{r^2}+\cfrac{Q_s^2(-M^2+Q_s^2+\cfrac{6}{m_2^2})}{3r^4}+\ldots  \\
B(r)^{-1} & = & 1-\cfrac{2M}{r}+\cfrac{Q_s^2}{r^2}+\cfrac{2Q_s^2(-M^2+Q_s^2+\cfrac{6}{m_2^2})}{3r^4}+\ldots 
\end{eqnarray}
where $Q_s$ is a scalar charge and $m_2$ is a massive spin-2 mode. This asymptote occurs far from the horizon. As the key point is to account the transfer of photons to photon orbit this asymptote appears to be applicable. Fig. \ref{r11} shows the dependence of the shadow size against scalar charge $Q_s$ for different values of $m_2$. The ends of the lines correspond to the effect described in \cite{Zakharov:2014lqa, Alexeyev:2019keq}: the big values of $Q_s$ and $1/m_2$ cause the absence of photon sphere. Decreasing of $m_2$ value causes the reduction of the shadow size. So additional observational data is required to constrain the model parameters. Limitations from \cite{2021} exclude only large values of $Q_s$ and $m_2$ ( Fig. \ref{r11}).

\subsection{Bumblebee model} 

The spherically symmetric solution in Bumblebee model has the form \cite{2018}:
\begin{eqnarray}\label{f91}
A(r) & = & (1-\cfrac{2M}{r})  \\
B(r) & = & \cfrac{1+l}{1-\cfrac{2M}{r}}
\end{eqnarray}
where $l=\xi b^2$, $\xi$ is the real coupling constant (with mass dimension equal to -1) which controls the non-minimal gravity-bumblebee interaction, $b^2=B^{\mu}B_{\mu}$.

Calculations show that the size of a shadow does not depend upon the parameter $l$ because $(1+l)$ can be put out of brackets and neglected. This effect occurs for all metrics where:
\begin{eqnarray}\label{f913}
A(r) & = & \Bar{A}(r)  \\
B(r) & = & \cfrac{1+l}{\Bar{B}(r)},
\end{eqnarray}
$l$ is a constant, $\Bar{B}(r)$ and $\Bar{A}(r)$ are the functions that can be presented as Tailor series. 

\begin{figure}
\begin{center}
\includegraphics[width=\linewidth]{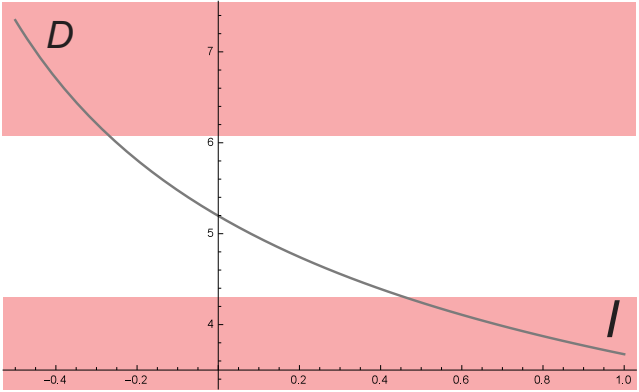}
\end{center}
\caption{The dependence of the shadow size $D$ upon parameter $l$ in alternative Bumblebee generalization with Schwarzschild approximation (in the units of $M$, $M=1$).}
\label{par_l_SH}
\end{figure}

The alternative generalization can be established as:
\begin{eqnarray}\label{f914}
A(r) & = & (1+l)\Bar{A}(r)  \\
B(r) & = & \cfrac{1}{\Bar{B}(r)},
\end{eqnarray}
The Schwarzschild metric can be used as the first approximation for $\Bar{B}(r)$ and $\Bar{A}(r)$ to examine such generalization:
\begin{eqnarray}\label{f9}
A(r) & = & (1+l)(1-\cfrac{2M}{r})  \\
B(r)^{-1} & = & 1-\cfrac{2M}{r}
\end{eqnarray}
The influence of the parameter $l$ on the size of the shadow is presented in Fig. \ref{par_l_SH}. For the other approximation of $\Bar{B}(r)$ and $\Bar{A}(r)$  the dependence has the same form. Setting the limits from M87 observation \cite{2021} on Schwarzschild approximation one obtains that $-0.3 < l < 0.45$.  

\subsection{$f(Q)$ Gravity} 
 
\begin{figure}
\begin{center}
\includegraphics[width=\linewidth]{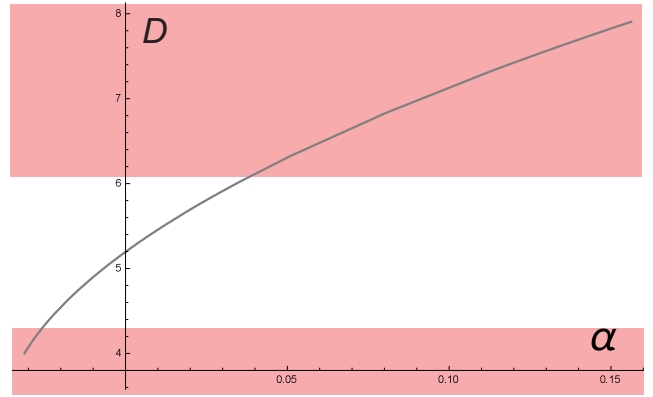}
\end{center}
\caption{The dependence of the shadow size $D$ upon parameter$\alpha$ in $f(Q)$ gravity in $M_{ren}$ units. }
\label{f(Q)1}
\end{figure}

$f(Q)$ gravity is a STEGR with non-metricity scalar $Q$ \cite{DAmbrosio:2021zpm}. We chose ($I^+$) approximate solutions beyond GR from constraints on the connection \cite{DAmbrosio:2021zpm000000000000000000000} in the form:
\begin{eqnarray}\label{f10}
A(r) & = & 1-\cfrac{2M_{ren}}{r}-\alpha\cfrac{32}{r^2} ,  \\
B(r)^{-1} & = & 1-\cfrac{2M_{ren}}{r}-\alpha\cfrac{96}{r^2} ,  \\
2M_{ren}  & = &  2M-\alpha(\cfrac{32}{3M}+c_1) ,
\end{eqnarray} 
where $a$ is a expansion parameter, $c_1$ is the integration constant, $M_{ren}$ is the re-normalized mass. Note that for a distant observer there is no difference between re-normalized and Schwarzschild masses. So we continue in $M_{ren}$ units. The influence of $a$ on the shadow size is shown in Fig. \ref{f(Q)1}. We set the limits from M87 observations \cite{2021} as: $-0.025 < \alpha < 0.04$

\subsection{Scalar Gauss-Bonnet gravity} 

The static spherically symmetric solution in scalar Gauss-Bonnet gravity was obtained analytically in \cite{PhysRevD.83.104002} and has the form: 
\begin{eqnarray}\label{GB1}
A & = & - f(r)[1 + \cfrac{\zeta}{3r^{3}f(r)}h(r)] ,\\
B & = & \cfrac{1}{f(r)}[1 - \cfrac{\zeta}{r^{3}f(r)}k(r)] ,
\end{eqnarray}
where
\begin{eqnarray}\label{GB2}
h(r):=1+\cfrac{26}{r}+\cfrac{66}{5r^2}+\cfrac{96}{5r^3}-\cfrac{80}{r^4}, \\
k(r):=1+\cfrac{1}{r}+\cfrac{52}{3r^2}+\cfrac{2}{r^3}+\cfrac{16}{5r^4}-\cfrac{368}{3r^5}, \\
f(r):=1-\cfrac{2}{r},
\end{eqnarray}
where $\zeta$ is the coupling parameter.

\begin{figure}
\begin{center}
\includegraphics[width=\linewidth]{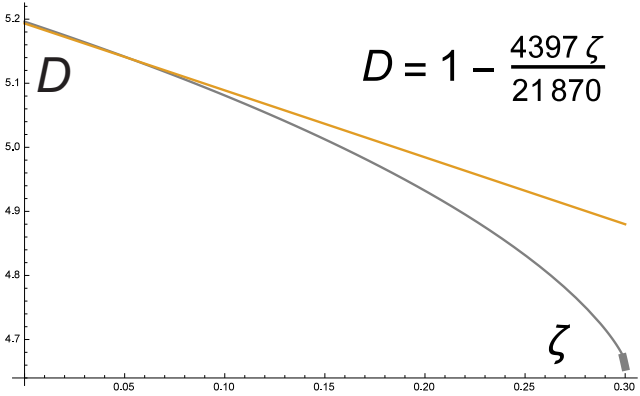}
\end{center}
\caption{The lower curve is the dependence of the shadow size  $D$ upon parameter $\zeta$ in scalar Gauss-Bonnet gravity (in the units of $M$, $M=1$). The top line is the first order approximation.}
\label{GB}
\end{figure}

In \cite{bauer2021spherical} dependencies of radius of photon sphere and shadow radius upon $\zeta$ are obtained in the first order relatively the coupling constant:
\begin{eqnarray}\label{GB5}
r_{ph}^{sGB} & = & 3[1-\cfrac{961}{2430}\zeta], \\
b_{c}^{sGB} & = & \sqrt{27}[1 - \cfrac{4397\zeta}{21870}] . 
\end{eqnarray}
We obtained numerically the solution with higher accuracy (Fig. \ref{GB}). There is no photon sphere when $\zeta > 0.3$. Analogously to \cite{Zakharov:2014lqa,Alexeyev:2019keq} this means that such an object has no shadow.

\section{Discussion and Conclusions}

The resolution of the first BH images from Event Horizon Telescope was approximately about half of the object's size \cite{Akiyama:2019fyp}. Further improving of ground-based equipment could increase the resolution only for few times  (not orders!). In addition the maximally possible size of the ground based network from radio-telescopes is already reached. As it was demonstrated earlier in \cite{Alexeyev:2019keq, Alexeyev:2020frp} and developed in previous sections the constraining of real extended gravity models requires few orders of the accuracy improving (not times!). Therefore the next step could be an orbiting telescope network. Moreover the measurement of the shadow size without additional data would be enough only for the models based on Reissner-Nordstrom metric. For theories with a more complicated BH space-time the amount of observational points must be increased. In a case of considering of the last stable orbit, strong gravitational lensing of bright stars \cite{Alexeyev:2019keq} and the distribution of shadow background intensity (as we showed above) the minimal resolution appears to be equal to $0.001$ of shadow size. This is the maximal assumption valid if the additional coefficients are compatible with the BH mass $M$. Therefore the study of the shadow from rapidly rotating object seems to be more perspective. With the same values of additional coefficients the necessary resolution would be about $0.01$ of shadow size \cite{Alexeyev:2020frp}. So, there is an additional reason to develop the Kerr-like BH shadows theory to explore the extended gravity models in astrophysics. 

We also calculate the dependencies of shadow size upon the model parameters in different extended gravity theories and set the limits on them using M87 BH observations. The results in Horndesky model with Gauss-Bonnet invariant, LQG, Bumbelby and Gauss-Bonnet scalar models lie in complete agreement with the M87* observations. For most of considered examples the model predictions are not pass the boundary established by the existing observational data. In addition in conformal gravity big values of $m_2$ and $Q_s$ must be excluded (for example if  $m_2=2$ then $Q_s<0.9$). In STEGR $f(Q)$ gravity M87 observations constrain $\alpha$ as $-0.025<\alpha<0.04$. In alternative Bumblebee generalization with Schwarzschild approximation one obtains that $-0.3 < l < 0.45$. These results demonstrates the maximum that could be distinguished when a BH rotation is not taken into account. 

Finally the approach without taking into account a BH rotation is valid only when a BHs rotation speed is small and can be neglected. When the rotation would be included into the consideration the amount of probes necessary to distinguish between different gravity models increases. ``As a compensation'' the requirements to the observational accuracy decrease. 

\section*{Acknowledgements}

This research has been supported by the Interdisciplinary Scientific and Educational School of Moscow University ``Fundamental and Applied Space Research''.





\begin{thebibliography}{10}
\expandafter\ifx\csname url\endcsname\relax
  \def\url#1{\texttt{#1}}\fi
\expandafter\ifx\csname urlprefix\endcsname\relax\def\urlprefix{URL }\fi
\expandafter\ifx\csname href\endcsname\relax
  \def\href#1#2{#2} \def\path#1{#1}\fi

\bibitem{Orosz:2011np}
J.~A. Orosz, J.~E. McClintock et~al., Astrophys. J. {\bf 742}, 84 (2011).

\bibitem{TheLIGOScientific:2016qqj}
B.~Abbott et~al.,  Phys.\ Rev.\ D {\bf 93} 12 (2016).

\bibitem{Abbott_2017}
B.~Abbott et~al.,  American Physical Society {\bf 119} 16 (2017).

\bibitem{Capozziello:2011et}
S.~Capozziello and M.~De~Laurentis, Phys. Rept. {\bf 509} 167 (2011).

\bibitem{dambrosio2021black}
F.~D'Ambrosio, S.~D.~B. Fell et~al., Phys. Rev. D {\bf 105} 2 (2021).


\bibitem{Sotiriou:2008rp}
T.~P. Sotiriou and V.~Faraoni, Rev. Mod. Phys. {\bf 82} 451 (2010).

\bibitem{DeFelice:2010aj}
A.~De~Felice and S.~Tsujikawa, Living Rev. Rel. {\bf 13} 3 (2010).

\bibitem{Charmousis:2011bf}
C.~Charmousis, E.~J. Copeland et~al., Phys. Rev. Lett. {\bf 108} 5 (2012).

\bibitem{Babichev:2017guv}
E.~Babichev, C.~Charmousis et~al., JCAP {\bf 4} 27 (2017).

\bibitem{Pfeifer:2021njm}
C.~Pfeifer and S.~Schuster, Universe {\bf 7~(5)} 153 (2021).

\bibitem{Mannheim_2011}
P.~D. Mannheim, Foundations of Physics {\bf 42~(3)} 388 (2011).

\bibitem{Myung_2019}
Y.~S. Myung and D.-C. Zou, Physical Review D {\bf 100~(6)} ( 2019).

\bibitem{casares2018review}
P.~A.~M. Casares, A review on loop quantum gravity (2018).
\newblock \href {http://arxiv.org/abs/1808.01252} {\path{arXiv:1808.01252}}.

\bibitem{De_Lorenzo_2015}
T.~De~Lorenzo, C.~Pacilio et~al., General Relativity and Gravitation {\bf 47~(4)} (Mar 2015).

\bibitem{Hu_2018}
J.-P. Hu, L.-L. Shi et~al., Astrophysics and Space Science {\bf 363~(10)} (Sep 2018).

\bibitem{PhysRevD.83.104002}
N.~Yunes and L.~C. Stein, Phys. Rev. D {\bf 83} 104002 (2011).

\bibitem{Horndeski:1974wa}
G.~W. Horndeski, Int. J. Theor. Phys. {\bf 10} 363 (1974).

\bibitem{Kobayashi:2019hrl}
T.~Kobayashi, Rept. Prog. Phys. {\bf 82~(8)} 086901 (2019).

\bibitem{2021ag}
Y.~Ageeva, P.~Petrov and V.~Rubakov, Physical Review D {\bf 104~(6)} (Sep 2021).

\bibitem{Barrau:2013ula}
A.~Barrau, T.~Cailleteau et~al., Class. Quant. Grav. {\bf 31} 053001 (2014).

\bibitem{Mannheim:2011ds}
P.~D. Mannheim, Found. Phys. {\bf 42} 388 (2012).

\bibitem{Arbuzov:2019rcl}
A.~B. Arbuzov and B.~N. Latosh, Universe {\bf 6~(1)} 12 (2020).

\bibitem{Alexeyev:2020lag}
S.~Alexeyev and D.~Krichevskiy, Phys. Part. Nucl. Lett. {\bf 18~(2)} 128 (2021).

\bibitem{Alexeyev:2021lav}
S.~Alexeyev, D.~Krichevskiy and B.~Latosh, Universe {\bf 7~(12)} 501 (2021).

\bibitem{2018}
R.~Casana, A.~Cavalcante, et~al., Physical Review D {\bf 97~(10)} (May 2018).

\bibitem{DAmbrosio:2021zpm}
F.~D'Ambrosio, S.~D.~B. Fell et~al., Phys. Rev. D {\bf 105~(2)} 024042 (2022).

\bibitem{Barrau:2003tk}
A.~Barrau, J.~Grain and S.~O. Alexeyev, Phys. Lett. B {\bf 584} 114 (2004).

\bibitem{Alexeyev:2019keq}
S.~Alexeyev, B.~Latosh, V.~Prokopov and E.~Emtsova, J.\ Exp.\ Theor.\ Phys. {\bf 128~(5)} 720 (2019).

\bibitem{Alexeyev:2020frp}
S.~Alexeyev and V.~Prokopov, J. Exp. Theor. Phys. {\bf 130~(5)} 666 (2020).

\bibitem{2021}
P.~Kocherlakota, L.~Rezzolla et~al., Physical Review D {\bf 103~(10)} (May 2021).

\bibitem{Dadhich:2000am}
N.~Dadhich, R.~Maartens et~al., Phys. Lett. B {\bf 487} 1 (2000).

\bibitem{Zakharov:2014lqa}
A.~F. Zakharov, Phys. Rev. D {\bf 90~(6)} 062007 (2014).

\bibitem{Pugliese:2010ps}
D.~Pugliese, H.~Quevedo and R.~Ruffini, Phys. Rev. D {\bf 83} 024021 (2011).

\bibitem{Bambi:2013nla}
C.~Bambi, Phys.\ Rev.\ D {\bf 87} 107501 (2013).

\bibitem{Shakura:1976xk}
N.~I. Shakura and R.~A. Sunyaev, Mon. Not. Roy. Astron. Soc. {\bf 175} 613 (1976).

\bibitem{Shaikh_2018}
R.~Shaikh, P.~Kocherlakota, et~al., Monthly Notices of the Royal Astronomical Society {\bf 482~(1)} 52 (2018).

\bibitem{Ageeva:2021yik}
Y.~Ageeva, P.~Petrov and V.~Rubakov, Phys. Rev. D {\bf 104~(6)} (2021).


\bibitem{bauer2021spherical}
A.~M. Bauer, A.~C\'ardenas-Avenda\~no et~al., Astrophys. J. {\bf 925~(2)} 119 (2022).

\bibitem{Akiyama:2019fyp}
K.~Akiyama, et~al., Astrophys. J. {\bf 875~(1)} L5 (2019).

\bibitem{Lima:2021las}
Lima, Junior., Haroldo C. D. et~al., Phys. Rev. D {\bf 8}, 084040 (2021).


\end{thebibliography}
\end{document}